\magnification=1000
\pretolerance=10000
\baselineskip=15pt

\centerline{\bf Strange stars in low-mass binary pulsar systems}

\bigskip
\bigskip
\centerline{J. E. Horvath}
\bigskip
\bigskip
\centerline{Instituto 
Astron\^omico e Geof{\'\i}sico, Universidade de S\~ao Paulo}

\centerline{Av. Miguel Stefano 4200 - S\~ao Paulo SP, 04301-904, Brazil}

\bigskip
\centerline{and}

\bigskip
\centerline{Steward Observatory, U. of Arizona}

\centerline{N. Cherry Av. 930, Tucson Arizona, USA}

\vskip 3 true cm
\centerline{\bf Abstract}
Based on observational facts and a variety of theoretical arguments we discuss 
in this work the possibility that pulsars in Low-Mass Binary Pulsar systems 
could be strange stars rather than neutron stars. It is shown that, although 
subject to reasonable uncertainties, the consideration of the physics of the 
SQM core and thin normal crusts leads to the prediction of several 
observed features of the magnetic field history of these systems 
whitin this working hypothesis.
\bigskip

\noindent
{Pacs 97.04.40.Dg}

\noindent
{Submitted to Int. Jour. Mod. Phys. D}
\vfill\eject

\noindent
{\bf 1. INTRODUCTION}
\bigskip

Low-Mass Binary Pulsars (hereafter LMBP) are generally regarded as 
important to probe the effects of substantial accretion rates on the 
structure and evolution of pulsars (see, for example, [1] for 
a review). In particular, it has become increasingly clear that 
the magnetic fields $B$ of isolated radio pulsars do {\it not} seem to decay 
on short timescales ($\tau_{decay} \, \sim \, 10^{7} \, yr$) as previously 
thought; while LMBP show strong evidence for $B$ decay. Accretion has been 
recently argued [2] to be involved in the very formation of at least 
one particular system ($PSR \, 1831-00$) 
and possibly {\it all} the pulsars in these systems by triggering 
an accretion-induced collapse (AIC) of a progenitor white dwarf; thus 
avoiding problems with formation 
in a type II supernova explosion [3].

On the other hand, it has long been speculated that, at least some pulsars 
(if not all [4-6]) must be strange stars if the 
strange matter (SQM) conjecture [4,7] is true; in the sense that 
hadronic matter is a metastable state decaying into a cold $u \, d \, s$ 
plasma (with a net gain of energy) under favorable conditions.

Which exactly are these "favorable conditions" is a matter of some 
controversy. While the idea of a mixed (neutron $+$ strange) population 
has been implicit in the literature, there is actually a fine-tunning 
problem to be explained, namely why a metastable state such as neutron 
matter can outlive a huge macroscopic time without decaying (we have 
argued elsewhere that the "natural" conversion is driven by the 
Kelvin-Helmholtz timescale $\sim \, 1 \, s$ after the formation of a 
protoneutron star [8]). On the other hand, it has been argued 
on observational grounds [9] that field radio pulsars 
{\it can not} be identified with strange stars; a strong conclusion that 
prompted refined calculations of "classical" strange stars [10] 
and exotic pulsar models [11]. Nevertheless, it should be 
acknowledged that the widely-spread notion that a neutron star must be 
highly compressed to reach the $neutron \, \rightarrow \, strange \, matter$ 
threshold at its center finds a natural setting in binary LMBP where 
AIC has formed the secondary before the end of the mass-transfer 
regime.

The purpose of this work is to point out that the identification of 
pulsars in LMBP systems with strange stars may allow the construction of a 
consistent scenario for their magnetic, thermal and evolutionary histories. 
Moreover, the physical features for the model to work seem to emerge quite 
naturally from simple estimates within reasonable uncertainties.

\bigskip
\noindent
{\bf 2. SCHEMATIC OVERVIEW OF A STRANGE STAR STRUCTURE AND EVOLUTION}
\bigskip

As is well-known, the structure of a strange star [5,10] consists 
of a dense degenerate SQM core sourrounded by a normal matter crust.
Both the Baym-Pethick-Sutherland [12] equation of state 
(which is generally considered as an accurate description for this 
density regime) and the case of an accretion-generated crust [13] (which is 
probably more adequate for the system we are considering) can be 
 parametrized by a polytropic expression 
$P \, = \, K \, \rho^{\Gamma}$ with sufficient accuracy. 
For the sake of the argument, we may properly describe the structure 
of this crust by integrating the hydrostatic equilibrium equation in 
the approximation $M \, = \, constant$ (compare, for example, with 
[10]) and get

$$
\delta R \, = \, \xi {\Gamma \over{\Gamma - 1}} 
{K R_{ss}^{2} \over{G \, M_{ss}}} \, \rho_{B}^{\Gamma -1}
\eqno(1)$$

where $R_{ss}$ is the strange core radius ($\sim$ star radius for 
$R_{ss} \, \gg \, \delta R$), $M_{ss}$ is the mass of the strange star 
in the same approximation, $\rho_{B}$ is the density at the base of 
the normal crust and $\xi \, \sim \, 0.65$ is a relativistic correction. 
This approximate (yet accurate) expression for the this crust allows us 
to relate all the relevant quantities to $\rho_{B}$ in a simple and 
useful way as will be explained below. We shall 
further assume that the crust forms on short timescales compared to 
the thermal evolution of the star. 

It is widely agreed that, due to the very features of the SQM core, 
the density of the normal matter at the base of the crust $\rho_{B}$ 
has to be limited by the neutron drip value 
$\rho_{D} \, \simeq \, 4.3 \, \times \, 10^{11} \, g \, cm^{-3}$. 
However, there is no obvious reason for the latter being the actual 
value, and any density lower than $\rho_{D}$ is, in principle, possible.
In fact, the authors of Ref.[14] have argued that 
$\sim \, \rho_{D}/5$ is the maximum value allowed by considering the 
mechanical equilibrium of the crust. In the first case, the exact 
accretion history of the star would be irrelevant, since after the 
condition $\rho_{B} = \rho_{D}$ is achieved, the dripped neutrons 
would be swallowed by the SQM core and we will always deal with a 
maximal crust. On the other hand, if the limiting density is 
$< \, \rho_{D}$, the mass of the crust 
$\delta M $ ($\leq \, 
10^{-6} \, M_{\odot}$ in this approximation) 
may depend on the history of the object 
(in the former case, the dominating isotope would be $^{118}Kr$, 
while in the latter we would find $^{80}Zn$ at the base of the 
crust). Fortunately, since $\Gamma$ is near $4/3$, the results 
will be quite insensitive to the actual value of $\rho_{B}$ 
provided the crust is not extremely tiny. Keeping this in mind, 
we shall leave the scaling explicit to a fiducial value 
$\rho_{0} \, = \, 10^{11} g \, cm^{-3}$ to allow for a range of 
possibilities. The maximal, neutron drip-limited case is obtained 
by setting $\rho = 4.3 \rho_{0}$, $Z = 36$, $X = 0.3$ below.

The thermal and magnetic history of 
a strange star will be very different according to the actual state of the 
SQM core. It has been longly recognized [15,16] that it is 
entirely possible that the quark liquid forms a superconducting state. 
The first schematic model calculations of compact stars 
cooling with superconducting SQM cores [17] 
have been recently refined and 
improved [18], to confirm that dramatic effects due to 
quark pairing can completely alter the quick-cooling signature of 
strange stars (see, for example, [19]). Since the critical 
temperature for pairing to occurr is very high in these models
($T_{c} \, \sim \, 0.1 \, MeV$, see [15,16]); we can neglect 
the very short time after the AIC formation event in which the SQM core 
remains in the normal state. Once it becomes superconducting the magnetic 
field $B$ should obey the diffusion equation 

$$
\nabla^{2} B \,  = \, {1 \over{D}} \, { {\partial B} \over {\partial t} }
\eqno(2)$$

\noindent
where $D \, = \, c^{2}/ 4 \, \pi \, \sigma_{sqm}$ is the diffusion 
coefficient. Acording to Refs.[20], the SQM conductivity $\sigma_{sqm}$
 can be expressed as 

$$
\sigma_{sqm} \, \simeq \, 10^{19}  
{\biggl( {\alpha_{c} T_{9}} \biggr)}^{-5/3}  
{\biggl( {\mu \over{300 \, MeV}} \biggr)}^{8/3} \, s^{-1}
\eqno(3)$$

\noindent
where $\alpha_{c}$ is the strong coupling constant, $T_{9} \, \equiv \, 
T/10^{9} \, K$ is the internal temperature and $\mu$ is the 
quark chemical potential. The consequence of such a  
conductivity is immediately clear: unless the estimate (valid for normal 
SQM)  happens to be wrong by many orders of magnitude, the magnetic flux 
is expulsed in a timescale as short as  
$\tau_{exp} \, \simeq \, R^{2} / D \, \simeq \, 3 \, \times \, 10^{3} \, yr 
(\mu / 300 \, MeV)^{8/3} \, (\alpha_{c} T_{9})^{-5/3}$; an astronomically 
small value [21]. Now, the conservation of the expulsed flux demands that 
the final value of the normal crust grows by a factor 
$B^{crust} \, = \, B^{core} {(R / {\delta R})}$ but, because of elastic 
stresses in the crust can not support magnetic stresses, the $B^{crust}$ 
must be limited [22] by 
$B^{crust}_{max} \, = \, (8 \pi \mu \Theta \delta R/ R)^{1/2}$, where 
$\mu$ is the lattice shear modulus and $\Theta \, \sim \, 10^{-2}$ is 
the shear angle. If we impose $\mu \, \sim \, 10^{26} \, 
dyn \, cm^{-2}$ for the former at $\rho \, = \, \rho_{0}$ then 
$B^{crust}_{max} \, \simeq \, 8 \, \times \, 10^{11} \, G$, 
which is the initial value 
expected for the field in a young BP (see below). Because its 
limited resistance, the crust would 
be blown off if the initial $B^{core}$ exceeds the threshold 
$B^{core}_{\ast} \, = \, 8 \, \times \, 10^{11} \, G \, 
(\delta R / R) \, \sim \, 10^{10} \, G$.  
We suggest that we may in fact be observing those LMBP in which 
the initial $B^{core}$ (for which we do not have any reliable information 
at all) remained lower than $B^{core}_{\ast}$. It is remarkable that 
none of the 24 known systems posesses a $B$ larger than the expected in 
the model.

\bigskip
\noindent
{\bf 3. FURTHER EVOLUTION OF THE MAGNETIC FIELD}
\bigskip

Due to ohmic currents the field, now totally confined to the crust if 
a small skin depth is neglected, should proceed to decay according to 

$$
{\partial B \over{\partial t}} \, = \, - \, {c^{2} \over{4 \, \pi}} \, 
\nabla \, \times \, 
{\biggl( {1 \over{\sigma_{crust}}} \, \nabla \, \times \, B \biggr)} \, \, .
\eqno(4)$$

 The solutions which are relevant to our problem are those satisfying 
$B \, = \, 0$ immediately above $\rho \, = \, \rho_{B}$; in agreement 
with the complete core flux expulsion expectation. Furthermore, since 
as discussed above, 
strange stars with crust satisfy the condition $\delta R / R \, \ll \, 1$, 
we may properly use the "thin crust approximation" [23] to obtain 
the decay timescale of the longest-lived dipole mode

$$
\tau_{d} \, = \, {2 \pi \over{c^{2}}} \xi {\Gamma \over{\Gamma - 1}} {\bar 
\sigma_{crust}} 
{K R_{ss}^{3} \over{G M_{ss}}} \, \rho_{B}^{\Gamma - 1} \, .
\eqno(5)$$

\noindent
where ${\bar \sigma}_{crust}$ is the average of the electrical conductivity 
over the crust. Since the conductivity will be dominated by the 
densest matter [23], an upper limit is set by replacing the average 
of the conductivity by its value at the drip point (that is 
${\bar \sigma}_{crust} \, \equiv \, \sigma_{crust} (\rho \, = \, \rho_{B})$)
hereafter.

As in most field decay models by ohmic dissipation, it is the value of 
$\sigma_{crust}$ which controls the behavior of $\tau_{d}$. At the higher 
temperatures $\sigma_{crust}$ is dominated by the {\it Umklapp} processes 
[24,25] and reads

$$
\sigma_{U} \, = \, 1.43 \, \times \, 10^{21} \, 
{\biggl( {\rho \over{\rho_{0}}} \biggr)}^{7/6} \, 
{\biggl( {X \over{0.375}} \biggr)}^{5/3} \, T_{9}^{-2} \, s^{-1} \, \, , 
\eqno(6)$$

\noindent
where the fraction of protons per nucleus $X$ has been scaled to its 
expected value at $\rho_{0}$ 
and $T_{9} \, \equiv \, T / 10^{9} \, K$ 
identified with the (isothermal) core temperature. Below a freezeout 
temperature $T_{F}$ the {\it Umklapp} processes are frozen and the 
conductivity is dominated by the impurity concentration $Q$. This 
$T_{F}$ can be expressed as 
 
$$
T_{F} \, = \, 1.73 \, \times \, 10^{7} \, 
{\biggl( {\rho \over{\rho_{0}}} \biggr)}^{1/2} \, 
{\biggl( {Z \over{30}} \biggr)}^{1/2} \, 
{\biggl( {X \over{0.375}} \biggr)} \, K \, \, ,
\eqno(7)$$

\noindent
where $Z$ is the charge of the dominant isotope $^{80}Zn$ at $\rho_{0}$.
Below this temperature the dominant conductivity due to impurity 
concentration takes the form [25]
 
$$
\sigma_{I} \, = \, 4.13 \, \times \, 10^{24} \, 
{\biggl( {\rho \over{\rho_{B}}} {X\over{0.375}} \biggr)}^{1/3} \, 
{\biggl( {Z \over{30}} \biggr)} \,  
{\biggl( {1 \over{Q}} \biggr)} s^{-1} \, \, ,
\eqno(8)$$

\noindent
where $Q \, \geq \, 10^{-3}$ is the mean square deviation of $A$ from 
its average [26] (it can be readily checked that, unless 
$Q$ is unexpectedly large, $\sigma_{I}$ does not dominate $\sigma_{U}$ 
for temperatures $T \, > \, T_{F}$ and we shall dismiss this possibility 
in the remaining of this work).

An important difference between these regimes is that while 
$\sigma_{U} \, \propto \, T^{-2}$, $\sigma_{I}$ does not depend on 
the temperature. The crust field decay is therefore different for 
$T \, > \, T_{F}$ than for $T \, < \, T_{F}$. In the first case the 
decay proceeds according to [23]
 
$$
B^{crust}(t) \, = \, B^{crust}_{max} \, \exp \, 
{\biggl( - {\int_{0}^{t} \, {d t' \over{\tau (t')}}} \biggr)} \, \, ,
\eqno(9)$$

\noindent
where 

$$
\tau \, = \,  7.2 \, \times \, 10^{3} \, 
{\biggl( {R \over{10 \, km}} \biggr)}^{3} \,
{\biggl( {1.4 \, M_{\odot} \over{M_{ss}}} \biggr)} \,
{\biggl( {\rho \over{\rho_{0}}} \biggr)}^{\Gamma + {1 \over{6}}}  
\times \, {\biggl( {X \over{0.375}} \biggr)}^{3} \, T_{9}^{-2} \, yr \, ,
\eqno(10)$$

\noindent
depends on time through the 
temperature $T$. In the second case, the decay is a simple 
exponential 

$$
B^{crust}(t) \, = \, B^{crust}_{F} \, \exp (-{t / {\tau_{I}}}) \, \, ,
\eqno(11)$$

\noindent
with $B^{crust}_{F}$ the value of the field at the end of the first regime 
and the time constant is 

$$
\tau_{I} \, = \, 2 \, \times \, 10^{7} \, (R / 10 \, km)^{3} \, 
\, (X / 0.375)^{5/3} 
\times \, (\rho / \rho_{0})^{\Gamma - {2 \over{3}}} \, 
(Z / 30) \, (1 / Q) \, yr \, , 
\eqno(12)$$

\noindent
which happens to be always $\geq \, 10^{9} \, yr$ for the expected parameters 
unless $Q$ happens to be very high.

A determination of the evolution of $B^{crust}(t)$ for $T \, > \, T_{F}$ 
requires the knowledge of the thermal history $T(t)$. The important 
point here is that SQM superconductivity renders a plateau in the 
$T$ vs. age curve [17,18] (which is absent in the case of a 
normal SQM core). The boldest approximation is to set $T \, = \, constant$ 
for the plateau era and use the relationship between the surface 
temperature and core temperature given in Ref.[27]; namely 
$T_{s} \, \simeq \, 10^{6} (T / 10^{8} \, K)^{0.55} \, K $ (which has been 
argued to be valid for strange stars as well [28]) to yield 
the crust field at any time $t \, \leq \, t_{pl}$

$$
B^{crust}(t) \, = \, B^{crust}_{max} \, \exp \, 
{\biggl[ {- {1 \over{7.2}} {\biggl( {t \over{10^{5} \, yr}} \biggr)} \, 
{\biggl( {<T_{s}> \over{10^{6} \, K}} \biggr)}^{3.64}} \biggr]} \, \, ,
\eqno(13)$$

\noindent
where $<T_{s}>$ is the average of the surface temperature in the 
plateau era. Since, according to recent calculations [18], 
$<T_{s}> \, \simeq \,  2 \, \times \, 10^{6} \, K$, 
we conclude that the field should decay by a factor of $\geq \, 10^{2}$ 
along the plateau era lasting $few \, \times \, 10^{5} \, yr$. 
Using a more accurate fit to 
Ref.[18] results $T_{s} \, = \, 10^{6.5} \, (yr / t)^{0.05} \, K$ 
we find the refined estimate of the decay along this epoch

$$
B^{crust}_{F}  \, = 
\, B^{crust}_{max} \, \exp \, 
{\biggl[ {- 1.45 {\biggl( {t \over{10^{5} \, yr}}} \biggr)}^{0.82} 
\biggr]} \, \, ,
\eqno(14)$$

\noindent
which leads to essentially 
identical conclusions. 

\bigskip
\noindent
{\bf 4. DISCUSSION}
\bigskip

Based on the results of the former section, it is tempting to suggest that a
decay of the field by a factor of $\sim \, 10^{3}$ in the first 
$10^{5}-10^{6} \, yr$ is built-in by the physics of strange star 
crusts with 
superconducting SQM cores. The same line of reasoning shows that, 
unless the impurity concentration is very high 
($Q \, > \, 10^{-2}$), 
further decay of $B^{crust}$ below $B^{crust}_{F} \, \sim \, 10^{9} \, G$ 
is inhibited and its value remains effectively frozen because of the 
larger value of the decay constant $\geq \, 10^{9} \, yr$, again an 
effect controlled by the microphysics of the thin normal crust of a 
strange star. 

From the astrophysical point of view a model in which LMBPs form from a 
symbiotic system ($\sim \, 1 \, M_{\odot}$ low-mass giant $+$ white dwarf) 
is attractive since the latter are abundant in the galaxy. 
According to the accepted scenario (see for example [29]), 
as the non-degenerate star leaves 
the main sequence its radius increases until filling 
the critical lobe and mass transfer starts. Along this mass-transfer stage 
AIC of the white dwarf happens, and because of the 
$\sim \, 0.1 \, M_{\odot}$ energy loss, the binary temporarily 
detaches. When accretion resumes the collapse of the neutron star to 
a strange star should follow after the accretion of $\geq \, 0.1 \, M_{\odot}$ 
from the companion, which is sufficient to drive the conversion by compression. 
The SS is born hot but must cool very quickly below $T_{c}$, and therefore 
expulse the interior field as described (the latest work on quark 
pairing seems to suggest much larger gaps of $\sim \, 100 \, MeV$ with 
potential important effects on the cooling which have not been explored as yet, 
see for example [30]). 
After $\sim \, few \, \times \, 10^{5} \, yr$ this field would decay 
to the "bottom value" $B^{crust}_{F} \, \simeq \, 10^{8} \, G$ (eqs. 9-11). 
We suggest that systems like PSR 1718-19 and PSR 1831-00 are quite young 
and their evolution downwards in the $B_{s} - P_{orb}$ plane (Fig. 1 of 
Ref. [2]) may be measurable. It is also important to note that spin periods 
in the millisecond range are possible for strange stars in LMBP [31], 
in agreement with observations; while they would be prohibited in the case of 
a neutron composition because of r-mode instability [32].
 
Our estimations above may 
be helpful for an 
interpretation of why we should expect the field to decay 
differently on two different 
timescales, a point not easily made for neutron star models where the 
much denser crust behaves differently and complete flux expulsion 
remains controversial. We finally 
remark that, at least in principle, the identification is also consistent 
with the lack of glitches and substantial timing noise of pulsars in LMBP 
systems (which is naturally expected from strange stars [33]); and 
explains the sharp contrast with isolated field radio pulsars $B$ evolution, 
whose composition need {\it not} to be exotic because of formation arguments.

\bigskip
\noindent
{\bf 5. ACKNOWLEDGEMENTS}
\bigskip
We would like to acknowledge the financial support of the Brazilian 
Agencies FAPESP (S\~ao Paulo) and CNPq through 
several forms of grants. M.P.Allen, 
P.Benaglia, G.A.Romero 
and, particularly, J.A.de Freitas Pacheco are greatfully 
acknowledged for useful suggestions.

\bigskip
\noindent
{\bf 6. REFERENCES}

\noindent
[1] D.Bhattacharya and E.P.J.van den Heuvel, Phys.Rep. 203, 1 (1991).

\noindent
[2] E.P.J.van den Heuvel and O.Bitzaraki, Astron.Astrophys. 297, L41 (1995). 

\noindent
[3] D.J.Helfand, M.Ruderman and J.Shaham, Nature 304, 423 (1983).

\noindent
[4] E.Witten, Phys. Rev. D 30, 272 (1984).

\noindent
[5] C.Alcock, E.Farhi and A.V.Olinto, Astrophys.J. 310, 261 (1986) ; 
P.Haensel, J.Zudnik and R. Schaeffer, Astron. Astrophys. 160, 121 (1986).

\noindent
[6] O.G.Benvenuto, J.E.Horvath and H.Vucetich, Int.Jour.Mod.Phys. A 6, 4769 
(1991).

\noindent
[7] A.Bodmer, Phys. Rev. D 4, 1601 (1971), see also H. Terazawa INS Report 338
(INS, Univ. of Tokyo, 1979) and S. A. Chin and A.K. Kerman, Phys. Rev. Lett. 
43, 1292 (1979).

\noindent
[8] O.G.Benvenuto and J.E.Horvath, Phys. Rev. Lett. 63, 716 (1989).

\noindent
[9] M.A.Alpar, Phys. Rev. Lett. 58, 2152 (1987).

\noindent
[10] N.K.Glendenning and F.Weber, Astrophys.J. 400, 647 (1992).

\noindent
[11] O.G.Benvenuto, J.E.Horvath and H.Vucetich, Phys. Rev. Lett. 64, 713 (1990).

\noindent
[12] G.Baym, C.Pethick and P.Sutherland, Astrophys.J. 170, 299 (1971).

\noindent
[13] P. Haensel and J.L. Zdunik, Astron. Astrophys. 229, 117 (1990).

\noindent
[14] Y.F.Huang and T.Lu, Astron.Astrophys.325, 189 (1997).

\noindent
[15] D.Bailin and A.Love, Phys. Rep. 107, 325 (1984).

\noindent
[16] J.E.Horvath, O.G.Benvenuto and H.Vucetich, Mod. Phys. Lett. A 7, 995 
(1992).

\noindent
[17] J.E.Horvath, O.G.Benvenuto and H.Vucetich, Phys. Rev. D 44, 1147 (1991). 

\noindent
[18] C.Schaab, B.Hermann, F.Weber and M.K.Wiegel, Astrophys.J.Lett. 480, 111 
(1997).

\noindent
[19] P.Pizzochero, Phys. Rev. Lett. 66, 2425 (1991).

\noindent
[20] H. Heiselberg and C.J. Pethick, Phys. Rev. D 48, 2916 (1993), note 
that this refined calculations give a different behavior than, 
e.g. P.Haensel and A.J.Jerzak, Acta Phys. Pol. B 20, 141 (1989).

\noindent
[21] Bailin and Love {\it op. cit.} had in 
fact first speculated that very short expulsion timescales were possible.

\noindent 
[22] M.Ruderman, Astrophys.J. 382, 576 (1991).

\noindent
[23] C.Pethick and M.Sharling, Astrophys.J. 453, L29 (1995).

\noindent
[24] V.A.Urpin, Sov. Astron. 36, 393 (1992).

\noindent
[25] V.A.Urpin and D.G.Yakovlev, Sov. Astron. 24, 303 (1980). 

\noindent
[26] E.Flowers and M.A.Ruderman, Astrophys.J. 215, 302 (1977). 

\noindent
[27] E.H.Gudmunsson, C.J.Pethick and R.Epstein, Astrophys.J. 272, 286 (1983).

\noindent
[28] V.V.Usov, Astrophys.J.Lett. 481, L107 (1997).

\noindent
[29] M. de Kool and J. van Paradijs, Astron.Astrophys. 173, 279 (1987).

\noindent
[30] T. Schaefer and F. Wilczek, hep-ph/9906512

\noindent
[31] J. Madsen, Phys. Rev. Lett. 81, 3311 (1998).

\noindent
[32] N. Andersson, K. Kokkotas and B.F. Schutz, Astrophys. J 510, 846 (1999).

\noindent
[33] P.B. Jones, Mon. Not. R.A.S. 246, 364 (1990).
\bye